# On the Limitations of Provenance for Queries With Difference

*Yael Amsterdamer*
INRIA
and Tel Aviv University

*Daniel Deutch*
INRIA
and Ben Gurion University

*Val Tannen*
University of Pennsylvania

**Abstract**
The annotation of the results of database transformations was shown to be very effective for various applications. Until recently, most works in this context focused on *positive* query languages. The provenance *semirings* is a particular approach that was proven effective for these languages, and it was shown that when propagating provenance with semirings, the expected equivalence axioms of the corresponding query languages are satisfied. There have been several attempts to extend the framework to account for relational algebra queries with *difference*. We show here that these suggestions fail to satisfy some expected equivalence axioms (that in particular hold for queries on "standard" set and bag databases). Interestingly, we show that this is not a pitfall of these particular attempts, but rather *every* such attempt is bound to fail in satisfying these axioms, for some semirings. Finally, we show particular semirings for which an extension for supporting difference is (im)possible.

## 1 Introduction

The annotation of the results of database transformations with provenance information has quite a few applications [13, 5, 22, 20, 9, 10, 24, 21, 17, 18, 25, 23, 2]. Recent work [16, 12, 14] has proposed a framework of *semiring annotations*. The idea is that every tuple of the database is associated with an element of a semiring $K$, and to propagate the annotations through query evaluation. This means that query constructs (of some expressiveness) are associated with operations in the semiring. For instance, the semiring addition corresponds to alternative derivation of a tuple, and thus e.g. union of two relations corresponds to adding up the annotations of tuples appearing in both; multiplication corresponds to joint derivation, thus a tuple appearing in the result of relational join will have annotation which is the multiplication of annotations of the two tuples that were joined to obtain it.

An important feature that guides the research on semiring-based provenance is that of *algebraic uniformity*, that is, the propagation of provenance through query evaluation is defined using only the semiring operations addition and multiplication (and the constants 0 and 1). This uniformity is fundamental since it allows provenance management to work with abstract annotations for tuples (more concretely, the free semiring of polynomials), and specialize to any semiring with "concrete" annotations (for tuples multiplicity, access control levels, cost, etc. [16, 12, 14]) when this information becomes available. Using this perspective, the above papers have developed the framework for *positive* relational algebra (as well as for *positive* datalog, and positive NRC). They have shown that for these languages, the semiring interpretation *satisfies the expected axioms* (e.g. the axioms that hold for set and bag semantics), for *every choice of semiring*. Moreover, the semiring axioms are *forced* by those of the corresponding algebras [16].

To reach beyond positive queries one would like to deal with *relational difference*. Classical work on incomplete databases [19] already provides a solution for set semantics but dealing with both set and bag semantics is mentioned in [16, 12] as a puzzling open problem. A breakthrough was obtained by Geerts and Poggi [13] who also made connections with earlier work that we further exploit here. Other semantics have also emerged since [15, 4], and we consider them below. Each such work has made a particular choice of semantics for provenance-aware relational difference. These semantics are all different, and in particular induce different axioms of query equivalence. Unfortunately, as we show below, for each of these definitions, some "natural" axioms that are expected to hold, fail in general.

Consequently, we take here a different approach. Instead of suggesting a particular semantics for provenance-aware relational algebra with difference, *we formulate a desired (sub)set of query equivalence axioms that are expected to hold from any such semantics, and ask: can one extend the semiring framework (entailed by a subset of the axioms) to define a semantics for which* these axioms will hold, for every choice of semiring? The main result of this paper is that the answer to this question is no, for very useful semirings.

While this is a negative result, we believe that it is not the final word in the research on extending the provenance semiring framework to queries with difference. Our result indicates that, unlike the case of positive relational algebra, a one-size-fits-all semantics is unlikely to exist for queries with difference. One possible solution is to choose particular semirings that can be extended to account for difference, while satisfying the desired axioms. We demonstrate this for a particular application of provenance, namely access control.

The rest of this paper is organized as follows. In section 2 we recall the correspondence between the semiring axioms and those of the positive relational algebra. Then, in section 3 we study the introduction of the difference operation. We review related work in section 4 and conclude in Section 5.

## 2 The Positive Relational Algebra

Our starting point is the technique of algebraic representation of annotation propagation that was introduced in [16]. This technique begins by assuming that the space $K$ of annotations is equipped with two operations $+, \cdot$ and two constants $0, 1$ used to define a semantics for positive relational algebra (SPJU) on $K$-relations, i.e., relations whose tuples are annotated with elements from $K$.

To define annotated relations we use the named perspective of the relational model [1]. Fix a countably infinite *domain* $\mathbb{D}$ of values (constants). For any finite set $U$ of *attributes* a tuple is a function $t : U \to \mathbb{D}$ and we denote the set of all tuples by $\mathbb{D}^U$. Given $(K, +, \cdot, 0, 1)$, a $K$-*relation* (with schema $U$) is a function $R : \mathbb{D}^U \to K$ whose *support*, $\text{supp}(R) = \{t \mid R(t) \neq 0\}$ is *finite*. For a fixed set $U$ we denote by $K$-*Rel* (when $U$ is clear from the context) the set of $K$-relations with schema $U$. We use the notation $t|_U$ for the restriction of the tuple $t$ to the attributes of $U$. We can then define the semantics of every relational algebra operator on $K$-*Rel*. Due to lack of space we repeat only two of the definitions, referring the reader to [16] for the others.

**Union** If $R_i : \mathbb{D}^U \to K$, $i = 1, 2$ then $R_1 \cup R_2 : \mathbb{D}^U \to K$ is defined by $(R_1 \cup R_2)(t) = R_1(t) + R_2(t)$.

**Natural Join** If $R_i : \mathbb{D}^{U_i} \to K$, $i = 1, 2$ then $R_1 \bowtie R_2 : \mathbb{D}^{U_1 \cup U_2} \to K$ is defined by $(R_1 \bowtie R_2)(t_1) = R_1(t) \cdot R_2(t_2)$ where $t_1 = t|_{U_1}$ and $t_2 = t|_{U_2}$.

As stated in [16], requiring that this semantics satisfy the relational algebra identities in figure 1(a) is equivalent to $(K, +, \cdot, 0, 1)$ satisfying the equational axiomatization in figure 1(b), i.e., forming a specific algebraic

(I1) $R \cup (S \cup T) = (R \cup S) \cup T$

(I2) $R \cup \emptyset = R$

(I3) $R \cup S = S \cup R$

(I4) $R \bowtie (S \bowtie T)$

(I5) $R \bowtie 1\!\!1 = R$

(I6) $R \bowtie S = S \bowtie R$

(I7) $R \bowtie (S \cup T) = (R \bowtie S) \cup (R \bowtie T)$

(I8) $R \bowtie \emptyset = \emptyset$

(a)

(A1) $a + (b + c) = (a + b) + c$

(A2) $a + 0 = a$

(A3) $a + b = b + a$

(A4) $a \cdot (b \cdot c) = (a \cdot b) \cdot c$

(A5) $a \cdot 1 = a$

(A6) $a \cdot b = b \cdot a$

(A7) $a \cdot (b + c) = (a \cdot b) + (a \cdot c)$

(A8) $a \cdot 0 = 0$

(b)

Figure 1: $K$-**relational algebra identities and algebraic axiomatization for the space** $K$ **of annotations**

structure called a *commutative semiring*. The correspondence is very tight: for $n = 1, \ldots, 8$, $K$-relations satisfy identity I$n$ iff $(K, +, \cdot, 0, 1)$ satisfies axiom A$n$.

Why the relational algebra identities in figure 1(a)? We rely on two important cases, namely set and bag semantics, corresponding to the commutative semirings $(\mathbb{B}, \vee, \wedge, \bot, \top)$ and $(\mathbb{N}, +, \cdot, 0, 1)$ resp. The identities in figure 1(a) hold in both cases. A second argument is that many more relational algebra identities (omitted here) for projection and selection already follow from A1-A8.

## 3 Adding Relational Difference

Consider now the full relational algebra, i.e., the positive algebra we already dealt with together with the relational difference operator. A natural approach to propagating annotations through difference [13], is to add an algebraic operation $-$ to the semiring structure and to define

**Difference** If $R_1, R_2 : \mathbb{D}^U \to K$ then $R_1 - R_2 : \mathbb{D}^U \to K$ is defined by $(R_1 - R_2)(t) = R_1(t) - R_2(t)$.

Following the same approach, we now search for an equational axiomatization for $(K, +, \cdot, 0, 1, -)$.

(I9) $R - R = \emptyset$

(I10) $\emptyset - R = \emptyset$

(I11) $R \cup (S - R) = S \cup (R - S)$

(I12) $R - (S \cup T) = (R - S) - T$

(I13) $R \bowtie (S - T) = R \bowtie S - R \bowtie T$

(a)

(A9) $\quad a - a = 0$

(A10) $\quad 0 - a = 0$

(A11) $\quad a + (b - a) = b + (a - b)$

(A12) $\quad a - (b + c) = (a - b) - c$

(A13) $\quad a \cdot (b - c) = a \cdot b - a \cdot c$

(b)

Figure 2: **Extending the identities and axiomatization for relational algebra including difference**

Similarly to our treatment of the positive relational algebra, we consider additional axioms I9-I13 that hold for both set and bag semantics, involving relational difference. As before, these identities correspond to the equational axioms A9-A13 for $(K, +, \cdot, 0, 1, -)$, depicted in Figure 2(b). Again as before, we can state additional identities involving difference and projection or difference and selection, e.g., $\sigma_{P_1}(R - S) = \sigma_{P_1} R - \sigma_{P_1} S$; and again these follow from A9-A13.

Next we show that satisfying axioms A1-A12 is relatively "easy" but further satisfying A13 complicates matters considerably. Indeed, by a result of Bosbach [6], axioms A1-A12 characterize the notion of *monus-semiring* [13] (*m-semiring* for short) [1].

For a commutative semiring $(K, +, \cdot, 0, 1)$, the structure $(K, +, 0)$ is a commutative monoid. To define m-semirings we need the following.

**Definition 3.1** *Let $(K, +, 0)$ be a commutative monoid. Define*
$$a \leq b \quad \Leftrightarrow \exists c \; a + c = b$$
*When $\leq$ is an order relation it is called the **natural order** on $K$ and the monoid $K$ is said to be **naturally ordered**.*

Examples for naturally ordered commutative monoids are the natural numbers $(\mathbb{N}, +, 0)$ and the booleans $(\mathbb{B}, \vee, \bot)$, but not $(\mathbb{Z}, +, 0)$. The next proposition shows that in such monoids, axioms A9-A12 uniquely determine the $-$ operation.

**Proposition 3.2 ([6] via [3])** *Let $(K, +, 0)$ be a naturally ordered commutative monoid. For any binary operation $a-b$ on $K$ the following are equivalent*

(i) *For all $a, b$, $a-b$ is the smallest $c$ such that $a \leq b+c$.*

(ii) *For all $a, b, c$ we have $a-b \leq c$ iff $a \leq b+c$.*

(iii) *Axioms A9-A12 hold.*

[13] defines m-semirings as commutative semirings whose additive monoid is naturally ordered and satisfies condition (i) in Proposition 3.2. Therefore, in an m-semiring the $-$ operation is completely determined by the $+$ operation. Bosbach's characterization implies:

**Corollary 3.3** $(K, +, \cdot, 0, 1, -)$ *is an m-semiring iff A1-A12 hold.*

In the sequel, for ease of reading we will not distinguish between semirings and their extensions to m-semirings (when such extension is possible); when this extension is possible we simply say that a particular semiring *is* an m-semiring.

Of course, $(\mathbb{B}, \vee, \wedge, \bot, \top)$, $(\mathbb{N}, +, \cdot, 0, 1)$, and $(\mathbb{R}^+, +, \cdot, 0, 1)$ are all m-semirings (but $(\mathbb{Z}, +, \cdot, 0, 1)$ or $(\mathbb{R}, +, \cdot, 0, 1)$ are *not*). The semiring of provenance polynomials [16] $(\mathbb{N}[X], +, \cdot, 0, 1)$ is also an m-semiring (albeit it lacks the universal property that it enjoys among commutative semirings [13], see discussion in the last section). Any boolean algebra is an m-semiring, with $a - b = a \wedge \neg b$. Moreover, any complete distributive lattice is an m-semiring because

$$b + inf\{c | a \leq b + c\} = inf\{b + c | a \leq b + c\} \geq a$$

In particular, the fuzzy semiring $fuzz = ([0, 1], max, min, 0, 1)$ is an m-semiring. Finally, any *finite* distributive lattice is complete, hence completely distributive, hence an m-semiring. In particular, the following are also of interest: (1) the m-semiring of all positive boolean expressions over a set of variables X, $PosBool[X]$, (2) the three value logic $TVL$, and (3) the *security* semiring $\mathbb{S} = (\mathbb{S}, \min, \max, 0_s, 1_s)$ where $S$ is the ordered set $1_s < \mathsf{C} < \mathsf{S} < \mathsf{T} < 0_s$ whose elements have the following meaning when used as annotations: $1_s$: public ("always available"), $\mathsf{C}$: confidential, $\mathsf{S}$: secret, $\mathsf{T}$: top secret, and $0_s$ means "never available" [12].

Additional m-semirings of interest are the tropical semiring $\mathbb{T} = (\mathbb{N}^\infty, min, +, \infty, 0)$, why-provenance semiring [8] and Trio semiring [5] (In [14] Green shows that why and Trio provenance can be captured via semirings), and the boolean expressions semiring $Bool[X]$.

While almost all the semirings considered in conjunction with positive queries are m-semirings, satisfying the

---

[1] This was obviously known to the authors of [13]; in fact, their citation of [3] led us to Bosbach's work.

| ⊨ A13 | ⊭ A13 |
|---|---|
| $\mathbb{B}$ | $TVL$ |
| $\mathbb{S}'$ | $\mathbb{S}$ |
| $\mathbb{T}$ | $fuzz$ |
| $\mathbb{N}, \mathbb{R}^+$ | |
| Trio[X], Why(X) | |
| Bool[X] | PosBool[X] |
| $\mathbb{N}[X], \mathbb{B}[X]$ | |

axiom A13 is another story. Table 3 summarizes our results on satisfaction of A13 for the above m-semirings. We next prove this characterization for some of these semirings (the proofs for the rest use similar techniques and are omitted for lack of space).

**Proposition 3.4** *If $(K, +, \cdot, 0, 1, -)$ is a distributive lattice that is an m-semiring and has two elements $a, b$ s.t. $a > b$ and $(a - b) \cdot b \neq 0$ then A13 fails in K.*

**Proof.** Recall that in a distributive lattice $+$ corresponds to $max$ and $\cdot$ to $min$. Indeed, we obtain $(a - b) \cdot b \neq 0$ but $a \cdot b - b \cdot b = b - b = 0$

**Corollary 3.5** *A13 fails in the security m-semiring $\mathbb{S}$, the m-semiring of positive boolean expressions $PosBool[X]$, and the fuzzy m-semiring $fuzz$.*

**Proof.** [sketch] In the security m-semiring, $a - b = a$ if $a$ is less secure than $b$, and 0 otherwise. For $a = S$ and $b = T$ we obtain $a - b = S$ and $(a - b) \cdot b = T \neq 0$.

For PosBool[X], let $x, y, z \in X$ be three distinct variables and $a = x \vee y \vee z, b = x \vee y$. We obtain $a - b = z$, and $(a - b) \cdot b = z \wedge (x \vee y) \neq 0$.

For $fuzz$, $a - b = a$ if $a > b$ (and 0 otherwise). Any two values $a > b \neq 0$ satisfy the requirement.

The security semiring is of particular interest, but A13 does not hold there. One practical solution for this particular case is to work with an alternative, "good" semiring $\mathbb{S}' = (P(\mathbb{S} - \{0_\mathbb{S}\}), \cup, \cap, \phi, \mathbb{S} - \{0_\mathbb{S}\})$. The elements of $\mathbb{S}'$ are all subsets of security credentials, and the idea is that every tuple is annotated, explicitly, with the credentials of all users that are allowed to see it (an empty set has the interpretation of "never available"). In particular, it is easy to embed the annotations of $\mathbb{S}$ in $\mathbb{S}'$ - every element $s$ in $\mathbb{S}$ is mapped to a set of all elements that are greater or equal to $s$ according to the order relation on $\mathbb{S}$. We may now use *set difference* as difference operator, and can show that the obtained m-semiring satisfies A1-A13. Note that a downside here, is that the size of annotations in $\mathbb{S}'$ is greater than of those in $\mathbb{S}$.

## 4 Related Work

Provenance information has been extensively studied in the database literature. Different provenance management techniques are introduced in [11, 7, 8, 5], etc., and we discussed them above in the context of their semiring representation [14]. Several semantics of provenance-aware *difference* have been proposed. Our result shows that no semantics can satisfy all axioms A1-A13 above. We have already shown that the *monus* semantics [13] fails (in general) at A13; we next identify where other suggested semantics fail.

$\mathbb{Z}$ **semantics [15].** In [15] the authors suggest a semantics for difference on $\mathbb{Z}$-relations, i.e. relations annotated by integers. In a nutshell, the semantics defines the annotation of a tuple $t$ in the result of relational difference $R - S$ to be its annotation in $S$ subtracted from its annotation in $R$; the resulting annotation may be negative. This definition fails to satisfy axioms A10 and A11.

**Semantics based on aggregate queries [4].** In [4] we have suggested a semantics for queries with *aggregation*, since nested aggregation queries can encode queries with difference. Consequently we have obtained a semantics for difference. Intuitively, our definition entails that a tuple $t$ appears in $R - S$ if it appears in $R$, but does not appear in $S$. When the tuple appears in the result of $R - S$, it carries its original annotation from $R$. I.e. the existence of $t$ in $S$ is used as a boolean condition. We can observe that this semantics fails to satisfy axiom A11, and the other axioms hold.

Of course, no semantics is necessarily "better" than other, and the choice of semantics thus depends on the application and on the desired axioms.

## 5 Conclusions and further work

The provenance polynomials semiring $\mathbb{N}[X]$ was shown in [16] to serve as "universal" provenance annotation domain for the positive relational algebra queries. [13] introduced m-semirings to deal with relational difference and noted that while $\mathbb{N}[X]$ is also an m-semiring, it lacks the universality property. Since m-semirings form an equational variety (axiomatized by A1-A12), [13] proposes to take the free m-semiring, for which there is a standard algebraic construction, as the "new" $\mathbb{N}[X]$.

We have identified an important and useful algebraic identity, A13, that fails for several important semirings who are, nonetheless, m-semirings. This leads us, for example, to search for alternative semirings for security.

It also follows that the free m-semiring fails to satisfy A13. By the way, $\mathbb{N}[X]$ does satisfy A13 but it still lacks the desired universal property. It seems that one should therefore take the free algebraic structure satisfying A1-A13 as the universal provenance annotation domain for the full relational algebra. However, the standard construction of this structure is awfully uninformative. A task for the future would be the study of this structure with the aim of providing more manageable and illuminating characterizations.